\documentstyle[12pt]{article}

\newcommand{\gfm}{ {\rm GeV/fm}^3}
\newcommand{\beq}{\begin{equation}}
\newcommand{\eeq}[1]{\label{#1} \end{equation}}

\newcommand{\lton}{\mathrel{\lower.9ex\hbox{$\stackrel{\displaystyle 
<}{\sim}$}}}
\newcommand{\ee}{\end{equation}} \newcommand{\ben}{\begin{enumerate}}
\newcommand{\een}{\end{enumerate}} \newcommand{\bit}{\begin{itemize}}
\newcommand{\eit}{\end{itemize}} \newcommand{\bc}{\begin{center}}
\newcommand{\ec}{\end{center}} \newcommand{\bea}{\begin{eqnarray}}
\newcommand{\eea}{\end{eqnarray}}
\newcommand{\beqar}{\begin{eqnarray}}
\newcommand{\eeqar}[1]{\label{#1} \end{eqnarray}}

\newcommand{\vx}{{\bf x}}
\newcommand{\vva}{{\bf v}_{\perp\alpha}}
\newcommand{\vxa}{{\bf x}_{\perp\alpha}}
\newcommand{\vxp}{{\bf x}_\perp}
\newcommand{\vp}{{\bf p}}
\newcommand{\vpa}{{\bf p}_{\perp\alpha}}

\newcommand{\pa}{p_{\perp\alpha}}

\begin{document}
\begin{flushright}
Columbia preprint CU--TP--707
\end{flushright}
\vspace*{1cm}
\setcounter{footnote}{1}
\begin{center}
{\Large\bf Hot Spots and Turbulent Initial Conditions\\ of Quark--Gluon Plasmas
in Nuclear Collisions\footnote{ This 
work was supported by the Director, Office of Energy
Research, Division of Nuclear Physics of the Office of High Energy and Nuclear
Physics of the U.S.\ Department of Energy under Contract No.\ 
DE-FG-02-93ER-40764.}}
\\[1cm]
Miklos Gyulassy, Dirk H.\ Rischke, and Bin Zhang\\ ~~ \\
{\small Physics Department, Pupin
Physics Laboratories, Columbia University} \\ 
{\small 538 W 120th Street, New
York, NY 10027, U.S.A.} \\ ~~ \\ ~~ \\
{\large \today}
\\[1cm]
\end{center}
\begin{abstract} 
As a result of multiple mini-jet production, initial
conditions of the QCD plasma formed in ultrarelativistic 
nuclear collisions may be inhomogeneous, 
with large fluctuations of the local energy density (hot
spots), and turbulent, with a chaotic initial transverse velocity field.
Assuming rapid local thermalization, the evolution of such  
plasmas is computed using longitudinal boost-invariant 3+1--dimensional 
hydrodynamics. We compare the evolution in case that the speed of sound in the 
plasma is constant to one resulting from an equation of state involving a 
strong first order transition, with a minimum of the velocity of sound 
as a function of energy density.  
We find that azimuthally asymmetric fluctuations and
correlations of the transverse energy flow, ${\rm d}E_\perp/{\rm d}y
{\rm d} \phi$, can develop in both cases due to the initial inhomogeneities.  
Hot spots also enhance significantly high--$k_\perp$ direct photon 
yields. In the case with a phase transition, the hadronization surface evolves
into an unusual foam-like structure.  Also in that case, we find that
hadronization is considerably delayed relative to the ideal gas case, just as
previous studies have found for homogeneous initial
conditions.  The time-delay signature of a rapid cross-over transition
region in the QCD
equation of state (as observable via meson interferometry) is thus found to
be remarkably robust to uncertainties in the initial conditions in heavy-ion
reactions.
\end{abstract}
\newpage

\section{Introduction}

At energies $\sqrt{s} > 100$ AGeV, copious mini-jet production in central
nuclear collisions is expected \cite{mini,hotglue,eskola} 
to be the main source of a very dense plasma of quarks and gluons with an
initial (proper) energy density an order
of magnitude above the deconfinement and chiral symmetry restoration
scale \cite{lattice}, $\epsilon_c \sim 1\; \gfm$ $(T_c\simeq 160$ MeV).
A large number of observable consequences of the formation of this new 
phase of matter
have been proposed based on a wide range of dynamical assumptions \cite{qm93},
and experiments are currently under construction to search for evidence of
that quark--gluon plasma (QGP) at the Relativistic Heavy--Ion Collider 
(RHIC) at Brookhaven.  
The observables include dilepton and hard direct photon yields,
high--$p_\perp$ jets and hadrons, strangeness and charmed hadron production,
identical meson pair interferometry, and collective transverse expansion.
Evidently, these and other proposed signatures depend sensitively on the
assumed ensemble of initial conditions as well as on the 
transport dynamics through the hadronization point.  

In this paper we explore the dependence of several observables on the
fluctuations of the initial conditions induced by  mini-jet production. 
Our central dynamical
assumption is that after a rapid thermalization time ($\sim 0.5$ fm/c) the
evolution of the plasma through hadronization can be approximated by
non-dissipative hydrodynamics.  We make this strong assumption to explore the
{\em maximal\/} effects that the sought after thermodynamic properties of
hot QCD matter may have on observables in ultrarelativistic
heavy-ion collisions.  Only
hydrodynamics gives a direct link between observables and the fundamental QCD
equation of state \cite{lattice}.  Finite dissipative effects generally tend to
reduce hydrodynamic signatures and require further knowledge of the transport
coefficients in the QGP as well.  Even in the ideal hydrodynamic
limit, however, the observables are sensitive to the initial 
formation physics of the plasma. 
It is this aspect of the problem that we concentrate on in this paper.

We show that in contrast to the conventional picture of QGP formation, the
initial mini-jet ensemble is characterized by a wide fluctuation spectrum
of the local energy density (hot spots) and of the collective flow field
(turbulence).  We also show that hydrodynamic evolution of that inhomogeneous,
turbulent ensemble can lead to novel observable consequences
including azimuthally asymmetric transverse energy fluctuations and 
enhanced radiance of hard
probes.  An especially significant result is the demonstration that the 
time-delay signature of the QCD phase transition, as 
found in previous studies
\cite{pratt,bertsch,risch_pipi} starting from homogeneous initial
conditions, survives this generalized class of more unfavourable
initial conditions.  Meson interferometry therefore remains one of the most
robust and generic probes of QGP formation given present uncertainties of the
initial conditions in heavy-ion collisions.

Before elaborating on the nature of the inhomogeneous, turbulent initial
conditions induced by the mini-jet production mechanism, we review first the
(homogeneous) ``hot-glue scenario'' \cite{hotglue,eskola} assumed in many
previous calculations of the observable consequences of QGP formation in $A+A$
collisions.  Mini jets are simply unresolved partons with 
moderate $p_\perp> 1$ GeV/c predicted from perturbative QCD.  
They are produced abundantly 
at collider energies because the inclusive cross section for gluon jets
with moderate $p_\perp > p_0=1\;(2)$ GeV/c 
rises to a value $\sigma_{jet}(p_0) \simeq 40\;(10)$
mb at $\sqrt{s}=200$ GeV, comparable to the total inelastic cross section.
Evidence for mini-jet production in $pp$ and $p\bar{p}$
reactions at collider energies has been inferred from the systematic rise with
increasing $\sqrt{s}$ of the yield of moderate--$p_\perp$ hadrons, 
of the central rapidity density, 
of the enhanced tails of multiplicity fluctuations, as well as
the flavour and multiplicity dependence of the 
mean transverse energy of hadrons (see Refs.\ \cite{wang}).

In the eikonal approximation to nuclear dynamics, the total number of mini-jet
gluons produced in central $A+A$ collisions is expected to be $A^{4/3}\sim
10^3$ times larger than in $pp$ collisions since each incident projectile
nucleon interacts with $\sim A^{1/3}$ target nucleons along the beam axis.
This simple geometric effect leads to a high rapidity density of mini-jet
gluons with ${\rm d}N_g/{\rm d}y \sim 300-600$ 
in $Au+Au$ as shown in Fig.\ 1a. The curves are calculated via the 
HIJING model \cite{wang} with shadowing and jet quenching
options turned off. Comparison of the curves for $p_0=1$ and 2 GeV/c gives an
indication of current theoretical uncertainties associated with the
extrapolation from $pp$ to $AA$ reactions.  The observed
$\sqrt{s}$ systematics of the $p\bar{p}$ data are best accounted for with
$p_0=2$ GeV/c in the HIJING model \cite{wang}. That model combines an eikonal
multiple collision formalism with the PYTHIA algorithm \cite{pythia} to 
generate
exclusive hard pQCD processes with $p_\perp>p_0$ and a variant of the LUND/DPM
string phenomenology \cite{lund} to model hadronization and account for the
non-perturbative, low--$p_\perp$ beam-jet fragmentation.  Other parton cascade
Monte Carlo models, such as developed in Ref.\ \cite{geiger}, using different
structure functions and hard pQCD cutoff schemes, can  account for the
$p\bar{p}$ data using a somewhat lower $p_0\simeq 1.5$ GeV/c.  
Theoretically \cite{mini}, the
scale separating the perturbative QCD domain from the non-perturbative 
(beam-jet) domain may be as low as $p_0=1$ GeV/c, although no hadronization
phenomenology has yet been developed with such a low scale that
could account for the available data. Another source of 
moderate--$p_\perp$ gluons in
very heavy-ion reactions has recently been proposed based
on a semi-classical treatment of the non-Abelian Weizs\"acker--Williams gluon
fields \cite{mclerran}. The above uncertainties in the initial conditions 
on the parton level are seen in Fig.\ 1b to correspond to approximately 
a factor of two uncertainty of the transverse energy produced per unit rapidity
in central $Au+Au$ collisions at RHIC energies.

Figure 1c shows that the difference between the cases $p_0=1$ and 2 GeV/c 
in the
HIJING model is due to the production of approximately twice as many gluons in
the moderate $p_\perp<4$ GeV/c region for $p_0=1$ GeV/c. 
(The $p_\perp$--spectra extend to $p_\perp=0$ because of initial
and final state radiation associated with mini-jet production.)
This difference is significantly
smaller than the lowest order pQCD estimate would give because of the
unitarized eikonal formalism used in HIJING to calculate multiple collisions
and multiple jet production. For $p_0=1$ GeV/c the mini-jet cross section is
comparable to the inelastic cross section. Due to Glauber multiple collision
shadowing, the number of mini jets in that case must scale less rapidly 
than with the number of binary $pp$ collisions. 

In Figure 1d the hadronization mechanism of the mini-jet gluons via the
string fragmentation mechanism is found to approximately
double the final hadron transverse energy 
distribution relative to Fig.\ 1b.
This is due to the pedestal or ``string'' effect and persists up to LHC
energies in this model. The mini-jet gluons are represented as kinks in the
beam-jet strings, and those kinks effectively boost the produced hadrons in the
transverse direction. The difference between the mini-jet contribution 
(Fig.\ 1b) and the final hadronic transverse energy distribution is due
to the string model implementation of beam-jet fragmentation in 
HIJING. That component necessarily 
involves non-perturbative, low--$p_\perp$ multi-particle production
and is presently under experimental study via heavy-ion 
reactions at lower CERN/SPS
energies ($\sqrt{s}=20$ AGeV) \cite{qm93}. 
While the string model provides an adequate phenomenology
of beam-jet fragmentation at those energies, it is not obvious that it will
continue to do so at RHIC and LHC. This represents a significant source
of theoretical uncertainty in calculating RHIC initial conditions. 
We will assume in this study that the extrapolation of the beam-jet physics 
via string phenomenology as encoded in the 
HIJING model holds up to RHIC energies. Possible sources of
fluctuations of the beam-jet component due, for example,
to ``colour rope'' formation
have been explored in the past \cite{ehtamec,iwagyu,biro}. 
However, at collider energies, the consequences of fluctuations 
due to the dominant mini-jet component have not been considered previously
to our knowledge.

In the hot-glue scenario, the thermalization proper time is assumed to be a few
times the mini-jet formation time
$\tau_0=\hbar/ p_0 \sim 0.1$ fm/c (our units are $c=k_B=1$). 
In fact, the initial pQCD mini-jet $p_\perp$--distribution is not far from
thermal as can be seen in Fig.\ 1c, but it turns out
that the gluon and quark multiplicities are below chemical equilibrium. 
Inelastic multi-gluon
production processes are therefore essential to achieve rapid equilibration.
Recent progress on radiative transport in QCD plasmas \cite{xiong,GyuWa,doksh}
suggests that equilibration is possible on scales less than 1 fm/c.

Taking longitudinal boost-invariant expansion into account
and assuming a cylindrical distribution of matter, 
the proper energy density averaged over transverse coordinates
at proper time $\tau >\tau_0$ is given by
the  Bjorken formula \cite{Bjorken}:  
\beq
\bar{\epsilon}(\tau) \simeq
\frac{{\rm d}E_\perp}{{\rm d}y} \;\frac{1}{\tau\pi R^2} \; \; .
\eeq{bj}
For ${\rm d}E_\perp/{\rm d} y \simeq 1$ TeV from Fig.\ 1d,
and $R=7$ fm, this yields an order-of-magnitude estimate of
$\bar{\epsilon}(\tau)\simeq 65\, [0.1\,{\rm fm/c}\,/\tau]\; \gfm $.
If only the gluon degrees of freedom are equilibrated,
then the temperature of the
``hot glue'' at the thermalization time
is $T(\tau_{th})\simeq [\bar{\epsilon}(\tau_{th})/5.26]^{1/4}\simeq 555\,
(\tau_0/\tau_{th})^{1/4}$ MeV. The evolution
of the temperature after $\tau_{th}$ of course
depends on the equation of state as well as the
assumed viscosity \cite{danielewic}.

In this scenario, observables of the plasma phase such as thermal dileptons or
photons can be computed as discussed in \cite{therma}.
Transverse expansion can also be taken into account \cite{blaizot} as well as
more realistic equations of state with a rapid cross-over transition
region \cite{risch_pipi}. For transverse expansion,
the temperature field, $T(\tau,\vxp)$, acquires a dependence on the
transverse coordinates. In the hot-glue scenario
azimuthal symmetry is naturally assumed for collisions at
zero impact parameter.

Another implicit assumption of the hot-glue scenario
is that the {\em initial\/} fluid 4--velocity field,
$u^\mu(x)$, vanishes in the transverse direction and that
the initial flow field
only reflects the Bjorken longitudinal expansion, 
\beq
u^\mu(t,\vxp,z)=(t/\sqrt{t^2-z^2},0,0,z/\sqrt{t^2-z^2}) \; \; .
\eeq{ubj}
Transverse expansion is allowed to develop in the course
of subsequent evolution, but initially 
the plasma is assumed to be quiescent in the above sense.

In this paper, we call into question
the commonly assumed azimuthal symmetry and smooth radial
profile of $b=0$ collisions and the above quiescent form of the initial 
velocity field. In the next section, 
we show that the mini-jet mechanism does not
support those simplifying assumptions unless the 
beam-jet component is much larger than estimated with HIJING.
In Section 3, the hydrodynamic evolution of inhomogeneous,
turbulent initial conditions is calculated and the novel
type of azimuthally asymmetric transverse shock collectivity
is discussed. In Section 4 the robustness of the time-delay signature
associated with a phase transition is demonstrated.
In Section 5, the enhanced radiance of direct photons from
hot spots is estimated. A brief summary is finally presented
in Section 6.

\section{The Inhomogeneous, Turbulent Glue Scenario}

Inhomogeneities arise in nuclear collisions
as a result of fluctuations of the number of soft and hard QCD  interactions
per unit transverse area. Fluctuations of the soft beam-jet
component have been considered before \cite{ehtamec,iwagyu,biro}, 
but at collider energies a new source of fluctuations that
are induced by mini-jet production
is expected to become dominant. Both types of fluctuations are
strongly correlated as a function of transverse coordinates.
In this paper, however, we consider fluctuations arising from only
mini-jet production and 
treat the soft component as  smooth background component of the plasma.

Each nucleon in a central $Au + Au$
collision suffers approximately $ A^{1/3}\pm A^{1/6} \simeq 6\pm 2$ 
inelastic collisions. 
Therefore, there are $\sim 40$ binary collisions per $\sigma_{in} \simeq
4$ fm$^2$. At RHIC energies, however, only a fraction
$\sigma_{jet}/\sigma_{in} \simeq 1/4$ of those produce mini jets. 
The fluctuations of the mini-jet number density
are substantial because $A^{1/3}$ remains relatively
 small even  for the heaviest nuclei.
In principle, two--nucleon correlations in the initial nuclei could
reduce the above type of geometric fluctuations. However, the available
data on high--$E_\perp$ production in nuclear collisions at the SPS
indicates sizable fluctuations, even beyond the independent nucleon gas
approximation \cite{baym}. 

In addition to geometric sources of fluctuations, 
the broad transverse momentum spectrum of mini-jet gluons in Fig.\ 1c 
further enhances the fluctuations of the energy and momentum
deposited by mini jets per unit area. These two effects conspire to induce
large fluctuations of the initial energy and momentum density
at early times as we show below.

The spectrum of hot spots can be computed
from the HIJING event list of parton transverse and longitudinal momenta
$(\vpa, p_{z\alpha}= p_{\perp\alpha}\sinh y_\alpha)$, and their
longitudinal and transverse production coordinates, $(z_\alpha=0,\vxa$).
The production points are taken from the initial transverse coordinates
of the nucleon (diquark-quark string) in which the PYTHIA \cite{pythia}
subroutines of HIJING embed the gluons.

To simplify the problem we assume Bjorken boundary conditions
and neglect additional fluctuations along the longitudinal coordinate.
The longitudinal velocity is thus assumed to vanish at $z=0$.
With this simplification, the transverse coordinate dependence
of {\em local\/} energy and transverse momentum density in the $z=0$ plane
can then be estimated from
\beqar
\left(\begin{array}{c}
{\cal E} (\tau,\vxp,z=0) \\ 
{\bf M}_\perp (\tau,\vxp,z=0)
\end{array}\right)
= \sum_{\alpha}\left(\begin{array}{c}
1\\ \vva
\end{array}\right) 
\frac{\pa}{\tau}\; F(\tau \pa) 
\; \delta^{(2)}(\vxp-\vxa(\tau)) \;
 \delta(y_\alpha)
\; \; .\eeqar{xebj}
The above formula takes into account the free streaming of
gluons 
in the transverse direction via $\vxa(\tau)=\vxa+ \vva\tau $ where
$\vva=\vpa /p_\alpha $, as well as
ideal Bjorken longitudinal expansion.
The factor $F(\tau \pa)\equiv(1+[\hbar/(\tau \pa)]^2)^{-1}$ 
is an estimate of the formation probability \cite{GyuWa} of 
partons with zero rapidity. High--$p_\perp$ gluons are produced first and 
lower--$p_\perp$ gluons later according to the uncertainty principle.
Before $\tau\sim \hbar/\pa$, those $\pa$--components
of the radiation field and the evolving
transient Weizs\"acker--Williams field of the passing nuclei \cite{mclerran}
interfere strongly and cannot be treated as a 
kinetic gas of partons or a plasma. 
The exact form of $F$ is
not important except for extremely small times $\tau<0.5$ fm/c
prior to thermalization.

The above expression, when averaged over transverse coordinates,
reduces to the original Bjorken estimate, eq.\ (\ref{bj}) with 
$\langle {\cal E}(\tau)\rangle \simeq \bar{\epsilon}(\tau)$.
In addition, the averaged transverse momentum density 
and hence the flow field vanishes, 
$\langle {\bf M}_\perp(\tau)\rangle=0$, up to finite multiplicity fluctuations.

To study the transverse coordinate dependence of the
initial conditions given discreet mini-jet phase space coordinates, 
we must specify the transverse, $\Delta r_\perp$,
and longitudinal, $\Delta y$, resolution scales corresponding
to an elementary fluid cell. 
The densities, coarse-grained on that resolution scale,
are obtained from (\ref{xebj}) by the simple substitution
\beq
\delta^{(2)}(\vxp-\vxa(\tau))\; \delta(y_\alpha) \rightarrow
\frac{\Theta(\Delta y/2 - |y_{\alpha}|)
}{\Delta r_\perp^2 \, \Delta y}
 \prod_{i=x,y}
\Theta(\Delta r_\perp/2- |x_{i\alpha}(\tau)-x_i|)
\; \; .\eeq{smear}

The uncertainty principle limits how small $\Delta r_\perp > 
\hbar/\Delta p_\perp$
could be. However, in our dynamic problem, causality restricts how large
it can be. At a given proper time $\tau$, after the two highly contracted
nuclei pass through each other, the
local horizon for any gluon in the comoving $(y_\alpha=0)$ 
frame only has a radius
$c\tau$. Thus, at the thermalization time, $\tau_{th}$, each gluon can be
influenced by only a small neighbourhood of radius $c\tau_{th}\simeq 0.5$ fm 
of the plasma. In order to {\em minimize\/}
fluctuations, we will take the transverse resolution scale to be the
maximal causally connected diameter, $\Delta r_\perp=2c\tau_{th}\simeq 1$ fm.
Also, we take the rapidity width to be $\Delta y=1$,
since gluons with larger relative
rapidities tend to be produced too late (due to time dilation) 
to equilibrate with gluons with $y=0$.
Note that the number, $N(\tau)=[R/(c \tau)]^2$,  of causally disconnected
domains in a nuclear area, $\pi R^2$, is initially very large and even at the
onset of hadronization, $\tau_h\sim 3$ fm/c, several disconnected domains
remain.

In Figure 1d we saw that
at $\sqrt{s} = 200$ AGeV, the HIJING model predicts that approximately
half of the final produced transverse energy
arises from beam-jet string fragmentation.
As emphasized before, it is unknown whether the string model coupling
between the beam-jet fragmentation and mini-jet fragmentation, that accounts so
well for $p\bar{p}$ multi-particle data \cite{wang}, can be applied to 
collisions of heavy nuclei. Effective string degrees of freedom may be quenched
due to the colour conductivity of the plasma \cite{selik1}, 
but soft field fluctuations associated with beam-jet fragmentation are 
likely to persist and contribute a background source of low--$p_\perp$ quanta
at some level.
In the present study, we do {\em not\/} hadronize the
partons via this default JETSET string mechanism in HIJING but only use their
phase space density to compute the mini-jet contribution to the initial
energy--momentum tensor $T^{\mu\nu}(x)$ as above. Since our dynamical
assumption is the applicability of hydrodynamics,
we model the soft beam-jet component by a homogeneous 
background fluid. We estimate the energy density of that component
from the HIJING final hadronic ${\rm d}E_\perp/{\rm d}y$. One advantage
of the hydrodynamic approach is that
by treating the beam-jet component as a fluid,
we need not specify further details of its uncertain
microscopic nature. In our case, the main  function of that background
fluid is to {\em reduce\/} the energy
density fluctuations induced by the mini jets.
The transverse coordinate distribution of that
soft component is assumed to reflect the density of wounded nucleons,
i.e., it is taken proportional to $(1-[r_\perp/R]^2)^{1/2}$.

If we take into account
the relatively long formation time of soft, $\langle p_\perp\rangle \simeq 0.3$
GeV/c, partons, only about half of the 
${\rm d} E^{soft}_\perp/{\rm d} y \simeq 0.5 $ TeV from Fig.\ 1d
contributes to the soft background. In that case, this soft component adds
a relatively small, smooth contribution $\sim 5\;\gfm$
to the central energy density at $\tau_{th}=0.5$ fm/c.

In Figures 2a,b the initial energy and momentum density profile
of a typical central $Au+Au$ event is shown as a function of the transverse
coordinates in the $z=0$ plane. This corresponds to a HIJING event
with the default mini-jet cutoff scale $p_0=2$ GeV/c. The profile for
$p_0=1$ GeV/c looks similar except the energy density scale increases
by approximately a factor of two. 
The striking feature illustrated by the lego plot in Fig.\ 2a is
the existence of several prominent ``hot spots'' with ${\cal E} > 20\;\gfm$
separated by $\sim 4-5$ fm.
In this event the hottest spot reaches an energy density of about
$40\;\gfm $. Between the hot spots are cooler regions
with energy density down to the soft background scale of $\sim 5 \;\gfm$. 

The turbulent nature of the initial conditions is illustrated
in Fig.\ 2b. An arrow plot representation of the transverse
momentum field is shown. The highest momentum regions tend to coincide
with the regions of highest energy density. The initial transverse 
flow velocities are found to be distributed broadly up to 
$\sim 0.5\,c$. We note that turbulence here is not inherent to QGP evolution
since the Reynolds number of the QGP, ${\cal R}e \sim R/l \sim 10$, is
not large. Thus, laminar flow is not expected to break up into
complex vortex structures. In our case, the turbulence of the QGP
initial conditions is {\em induced\/} by the external mini-jet production
mechanism. This type of turbulence is analogous to flow induced
by the blades of a mixer in a bowl of liquid.  

In Figure 2c the distribution of energy densities is shown,
coarse-grained over 1 fm$^3$ cells and averaged over 200 HIJING events.
Only cells in the central region with $r_\perp<4$ fm 
are considered in this histogram to reduce fluctuations from the 
less interesting surface region. The event-averaged energy density
$\langle {\cal E}(\tau_{th})\rangle \simeq 12\; \gfm$ 
includes the soft background contribution discussed above.
However, the distribution is highly asymmetric
with relatively high probability of large and small fluctuations.  
The rms relative fluctuation of the initial energy density is found to be
$\Delta {\cal E}/\langle {\cal E} \rangle
\simeq 0.7$ for this assumed soft background level.

In Figure 2d the distribution of the effective local gluon temperature,
$T_{\rm eff}(\tau_{th},\vx_\perp)=({\cal E}(\tau_{th},\vx_\perp)/5.26)^{1/4}$, 
corresponding to Fig.\ 2c is shown. 
(This estimate of the temperature neglects collective flow velocities
and that the gluon number as computed in HIJING is not in chemical
equilibrium.)
The local temperature is seen to
fluctuate around the mean $\sim 350$ MeV with an rms width
of $\Delta T\simeq 60$ MeV. 
It is clear from the above that the mean values of ${\cal E}$ and
$T$ do not provide an adequate characterization of such an ensemble of
initial conditions.
The ensemble of mini-jet initial conditions has a  
broad fluctuation spectrum primarily 
because the local event horizon is so small at early times.
Fluctuations of ${\cal E}$ in finite volumes arise even in 
an ideal Stefan--Boltzmann thermal ensemble 
with $\langle \epsilon \rangle=KT^4$.
A simple estimate of the magnitude of thermal fluctuations
is given by $\Delta \epsilon / \langle \epsilon \rangle \simeq 
2/(K T^3 V)^{1/2} \simeq 0.5$ for $T=350$ MeV, $K\simeq 5.26$ 
(if only the gluon degrees
of freedom are assumed to be equilibrated), $V=4 (c\tau_{th})^3$
(for $\Delta y = 1$). This is comparable
to the fluctuations induced dynamically by the mini-jet formation.
The spectrum of fluctuations differs from an ideal
thermal one because they are driven here by Glauber nuclear geometry
and the pQCD mini-jet spectrum.

In Figure 3a the smooth, event-averaged energy density profile
is compared to the fluctuating energy density profiles of three
other separate HIJING events. The azimuthally symmetric, event-averaged 
surface corresponds to the one usually assumed
in the hot-glue scenario. The three individual events in parts b--d,
on the other hand, show that the dynamical
fluctuations extend up to 40 $\gfm$ (see Fig.\ 2c) and
cause strong deviations from the smooth, event-averaged profile.
The shaded contour plots above the surface plots provide another
representation of the azimuthally asymmetric inhomogeneities
caused by mini-jet production at the event-by-event level.

It is important to note that 
the hot spots are not due to isolated hard pQCD jets, 
but rather to the accidental coincidence
of several moderate--$p_\perp$ (see Fig.\ 1c) mini jets 
in the same 1 fm$^3$ volume. This is seen by comparing 
the gluon number density profiles in Fig.\ 4 to the
corresponding energy density profiles in Fig.\ 3.
The typical gluon density is seen to be about 5--10 fm$^{-3}$
at this early time $\tau_{th}=0.5$ fm/c. This includes, in addition
to mini jets, softer gluons from
initial and final state radiation as well as the soft partons from
the beam-jet fragmentation component.

Again we emphasize that Figs.\ 2--4 correspond to the
minimal fluctuations from the point of view of hydrodynamic
evolution. Only at later times, when the energy density is significantly
lower, can the fluctuations 
be significantly reduced by coarse graining on much larger resolution scales.
The dilemma is that {\em if\/} indeed the local thermalization is so
short as hoped for in the hot-glue scenario, then 
inhomogeneous and turbulent initial conditions must be considered
when calculating signatures of QGP formation.
On the other hand, if the thermalization time turns out to be much longer
than current estimates, then hydrodynamics is of course inapplicable and
the observables sensitive to the initial ``hot'' phase of the reaction
will not provide information
about the sought after {\em thermal\/} properties of the dense
QGP. For our purpose of exploring optimal signatures associated with
the thermal properties of dense matter, we must
rely on rapid thermalization and therefore live with 
inhomogeneities in calculating plasma observables. Inevitably this means
that many observables will depend strongly on the
precise form of the ensemble of initial conditions. The mini-jet 
initial conditions with a string model beam-jet background 
represent a particular ensemble obtained 
in extrapolating present $pp$ phenomenology to nuclear collisions
via the HIJING model. We will refer to the above ensemble of initial conditions
as the ``turbulent-glue scenario'' to contrast it with the simpler
initial conditions assumed in the hot-glue scenario.

Examples of plasma probes sensitive to the
highest temperature phase
are high-mass dilepton pairs, direct photons, and heavy quark
production. They are exponentially  sensitive to the
fluctuation spectrum of local temperatures because of 
the Boltzmann suppression factor, $\exp(-M_\perp/T)$.
The hot-glue and turbulent-glue scenarios differ considerably
in the predicted yields 
of such observables because the ensemble average of hydrodynamically
evolved turbulent initial conditions and the hydrodynamic evolution of
ensemble-averaged initial conditions are not equivalent.

\section{Hydrodynamic Evolution of the QGP}

\subsection{The SHASTA Algorithm and Tests}

In order to explore how such turbulent plasmas may evolve, we
solve numerically the hydrodynamic equations given the
above ensemble of  initial conditions.
Hydrodynamic evolution of homogeneous plasmas has already been studied 
extensively \cite{risch_pipi,blaizot}. In most previous studies, azimuthal
(cylindrical) symmetry was assumed to simplify the problem. 
One exception is the study
of homogeneous but azimuthally asymmetric initial conditions
produced in non-central nuclear collisions \cite{ollit}.
Our study focuses exclusively on $b=0$ collisions, where cylindrical symmetry
of {\em ensemble-averaged} observables must always hold.
At the event-by-event level, azimuthal symmetry is of course broken by the 
probabilistic nature of hadronic interactions. However, we are not 
interested in statistically uncorrelated fluctuations, rather only 
in the azimuthally asymmetric correlations that can evolve dynamically out of  
specific inhomogeneous mini-jet initial conditions.

We assume longitudinal boost-invariance, i.e.,
inhomogeneities in the rapidity dimension are neglected in this first study.
Mini jets from HIJING actually lead to hot-spot fluctuations 
that extend only one unit in rapidity.
The true 3--dimensional inhomogeneities are therefore significantly
larger than what will be studied here 
assuming that the fluid density is {\em constant\/} along
fixed proper time surfaces $\tau=\sqrt{t^2-z^2}$.
The full treatment of the problem will require
much more elaborate 3--dimensional simulations in the future. At present,
a 3--dimensional code is in the development phase.

A very important advantage of using a hydrodynamic
approach is that hadronization can be taken into account 
using the non-perturbative equation of
state, $p(\epsilon)$, as deduced from lattice QCD simulations.  
The price paid for hydrodynamics is of course the necessary assumption
that the equilibration rate is large compared to space--time gradients of
the fluid $T^{\mu\nu}$--field.

In order to compute the evolution of turbulent initial
conditions we solve the equations of relativistic hydrodynamics,
\beq
\partial_{\mu} T^{\mu \nu} =0\,\, .
\eeq{hyd}
We rely on the most recent optimistic estimates \cite{xiong,GyuWa,doksh}
that suggest that the local equilibration time in a QGP may be
short enough to neglect dissipative effects. In that case,
$T^{\mu \nu} = (\epsilon + p) u^{\mu} u^{\nu} - p g^{\mu \nu}$
is the energy--momentum tensor for an ideal fluid, where
$\epsilon$ is the (proper) energy density and $p$ the pressure in the 
local rest frame of a fluid element moving with 4--velocity $u^{\mu} = 
\gamma (1, {\bf v})$ in the computational frame (${\bf v}$ is the 
3--velocity, $\gamma = (1-{\bf v}^2)^{-1/2}$, and $g^{\mu \nu} = 
{\rm diag} (+,-,-,-)$ is the metric tensor). 
The equations (\ref{hyd}) are closed by specifying an
equation of state $p(\epsilon)$.

Since we assume boost-invariance in $z-$direction, it suffices
to solve the equations of motion at $z=0$. Furthermore, since
boost-invariance implies $v^z \equiv z/t$,
the four eqs.\ (\ref{hyd}) can be simplified to yield the three equations
(valid at $z=0$)
\bea
\partial_t\, {\cal E} + \partial_x\, [({\cal E}+p)v^x] + 
\partial_y\, [({\cal E}+p)v^y] & = & -\, F({\cal E},p,t) \, \, ,  \nonumber \\
\partial_t\, M^x + \partial_x\, (M^x v^x+p) + \partial_y \, (M^x v^y) & = & -\,
G(M^x,t) \, \, , \label{eomb} \\
\partial_t\, M^y + \partial_x\, (M^y v^x) + \partial_y \, (M^y v^y+p) & = & -\,
G(M^y,t) \, \, , \nonumber
\eea
where $F({\cal E},p,t) \equiv ({\cal E}+p)/t$, and $G(M,t) \equiv M/t$. 
(Our notation is $T^{00} \equiv {\cal E}$, $T^{0i} \equiv M^i, \; i=x,y$.)
These equations are solved numerically via a two--step operator splitting
method. The first operator splitting step decomposes the problem
into solving the above system of equations with
$F=G=0$ (this corresponds to purely two--dimensional fluid motion), and
then updating the obtained solution $\tilde{{\cal E}},\, \tilde{M}^x,\, 
\tilde{M}^y, \, \tilde{p}, ...$ according to the ordinary 
differential equations
\beq
\frac{{\rm d}{\cal E}}{{\rm d}t} = - F({\cal E},p,t)\,\,, \,\,\,\,
\frac{{\rm d}M^i}{{\rm d}t} = - G(M^i,t)\,\,, \;\; i=x,y\,\,,
\eeq{eomb2}
i.e., more explicitly one corrects
\beq
{\cal E} = \tilde{{\cal E}} -  F(\tilde{{\cal E}},\tilde{p},t)\, 
{\rm d}t \,\,, \,\,\,\,
M^i = \tilde{M}^i - G(\tilde{M}^i,t)\, {\rm d}t \,\,, \;\; i=x,y\,\,.
\eeq{eomb3}
This method was originally suggested by Sod \cite{sod} and was proven 
to be adequate for treating the analogous, 
azimuthally symmetric boost-invariant problem in \cite{risch_pipi}.

The solution to (\ref{eomb}) with $F=G=0$ itself involves another
operator splitting step, i.e., one first solves the system
of equations
\bea
\partial_t\, {\cal E} + \partial_x\, [({\cal E}+p)v^x] & = & 0 
\, \, , \nonumber \\
\partial_t\, M^x + \partial_x\, (M^x v^x +p) & = & 0 \, \, , \label{eomb4} \\
\partial_t\, M^y + \partial_x\, (M^y v^x)  & = & 0 \, \, , \nonumber
\eea
corresponding to transport of the hydrodynamic fields in $x-$direction
only, and with the solution to this set of equations one solves
\bea
\partial_t\, {\cal E} + \partial_y\, [({\cal E}+p)v^y] & = & 0 
\, \, , \nonumber \\
\partial_t\, M^x + \partial_y\, (M^x v^y) & = & 0 \, \, , \label{eomb5} \\
\partial_t\, M^y + \partial_y\, (M^y v^y+p)  & = & 0 \, \, , \nonumber
\eea
corresponding to transport in $y-$direction only.
Equations (\ref{eomb4}) and (\ref{eomb5}) are solved with the phoenical SHASTA 
algorithm \cite{SHASTA} 
as presented in \cite{test1}, with half-step updating of the source terms
and simplified source term treatment.
The transport steps (\ref{eomb4}) and (\ref{eomb5})
are alternated between successive time steps
to minimize systematical errors in the propagation. 
Per each time step, the
fields are propagated according to (\ref{eomb4}) and (\ref{eomb5}),
and finally corrected with (\ref{eomb3}).

The fluid evolution is studied for two idealized forms of the QGP equation
of state. One is the ideal relativistic gas case
\beq
p (\epsilon) =\epsilon/3 = K\, T^4/3 \; \; .
\eeq{eos_ideal}
Here, $K\simeq 5.26$, corresponding to an equilibrated gluon gas.
In the second case, a Bag model equation of state
with  a strong first order transition at a critical temperature
$T_c$ is assumed:
\beq
p(\epsilon)= \left\{ \begin{array}{cl}
(\epsilon- 4B)/3\;  & {\rm if } \; \epsilon>\epsilon_Q\; , \\
p_c \;  & {\rm if } \; \epsilon_Q \geq \epsilon \geq \epsilon_H\; , \\
\epsilon/3 \;  & {\rm if } \; \epsilon_H > \epsilon \; .
\end{array}
\right.
\eeq{bag}
With the Bag constant $B$ and ratio of effective plasma and hadronic
degrees of freedom $r=K_Q/K_H$ fixed,
the energy density of the mixed phase is bounded by 
\beqar
 \epsilon_Q &=& \frac{4r - 1}{r - 1} \; B \nonumber \\
        \epsilon_H &=& \frac{3}{r-1} \; B
\; \; .
\eeqar{eqh}
We take here $B=0.39691\;\gfm$ and  $r=37/3$.
For this choice, $\epsilon_Q \simeq 1.7 \; \gfm$ and 
$\epsilon_H \simeq 0.1 \; \gfm$, the critical pressure is
$p_c=\epsilon_H/3$ and the critical temperature is
$T_c\simeq 169$ MeV.
Hydrodynamics with a more realistic equation of state 
with a finite cross-over region $\Delta T/T_c \sim 0.1$ was considered in
Ref.\ \cite{risch_pipi}. In the present exploratory study, only the two 
idealized forms above will be contrasted.

The evolution equations are solved on a two--dimensional cartesian 
$200 \times 200$ mesh with grid spacing $\Delta x=0.2$ fm.
The Courant--Friedrichs--Lewy number is taken as
$\lambda \equiv \Delta t/\Delta x =0.4$.
The cartesian grid breaks the isotropy of space and might lead to
instabilities.
As a first check to determine 
whether our multi-dimensional algorithm tends to produce such
numerical artifacts, we consider the expansion of a cylindrically
symmetric Gaussian hot spot with radius 2 fm and peak energy 
density ${\cal E} = 30\; \gfm$ at rest. The time evolution should respect the 
initial cylindrical symmetry of the problem.
In Figure 5a we show the initial (calculational frame) energy density
profile ($T^{00} \equiv {\cal E}$). 
Figure 5b shows the energy density profile after evolving the
hydrodynamical equations with the standard version of the
SHASTA algorithm as described in \cite{test1} 
and the ideal gas equation of state (\ref{eos_ideal})
at time $t=14.9$ fm/c (for
the sake of clarity we show the profile for
positive $x, y$ only, the other quadrants are symmetric).
The observed strong fluctuations which break cylindrical symmetry are
due to the following:
the flux limiter in our version of the SHASTA 
(cf.\ eq.\ (18) of \cite{test1}) prevents the occurrence of 
unphysical extrema {\em only along\/} the direction of propagation, i.e., the
$x-$ and $y-$direction. Off the grid axis, small perturbations are
not smoothed out and can grow to produce the features seen in Fig.\ 5b.

One way to cure this is to use a multi-dimensional flux limiter as
proposed by Zalesak \cite{zalesak}. Here, however, we choose 
the simpler method of reducing the coefficient in front
of the antidiffusion fluxes (default value 1/8, cf.\ eq.\
(17) of \cite{test1}; the necessity to reduce this
coefficient occurs also in other, purely one--dimensional situations,
cf.\ the discussion in \cite{test1,test2}).
Figure 5d shows the same situation after evolving with the antidiffusion
coefficient reduced to 70\% of its default value, i.e., to 0.7/8=0.0875.
This obviously strongly reduces symmetry breaking. Although this 
prescription increases the numerical diffusion of the algorithm and thus
produces physical entropy, all our following results are generated with
this reduced antidiffusion in order to avoid spurious azimuthal
asymmetries of the transverse energy, ${\rm d} E_\perp/{\rm d}y {\rm d} \phi$,
of purely numerical origin. In Figure 5c we show the
energy density after evolving with the equation of state with the first order 
phase transition (\ref{bag}).
Also in this case, the reduction of the antidiffusion helps to preserve
the cylindrical symmetry of the problem.

Another test of the algorithm is to check how well it reproduces 
test cases with analytically known solutions.
One such test would be the expansion of a cylinder with a
sharp surface. Here, however, we focus instead on the more 
physically relevant problem of the expansion of the cylindrically symmetric 
(smooth) Gaussian studied in the previous figure. 
Although that problem is effectively
one--dimensional, it does not have a purely
analytical solution. We may, however, compare the numerical solution generated
with our multi-dimensional SHASTA algorithm to that
obtained with a one--dimensional algorithm which is known to reproduce
analytical solutions very well, namely the relativistic 
Harten--Lax--van Leer--Einfeldt (RHLLE) 
algorithm \cite{test1,test2,schneider,dhrmg},
for our purposes modified with a Sod operator splitting step 
as described in \cite{risch_pipi} 
to account for longitudinal boost-invariance (that operator splitting 
is in fact analogous to the one in eq.\ (\ref{eomb3})).

In Figure 6 we compare energy density profiles along the $x-$axis
for (a) the ideal gas equation of
state and (b) the Bag model equation of state with phase transition.
For the RHLLE run we used a 2000 cell grid with $\Delta x = 0.02/\lambda$ 
fm, $\lambda = 0.99$. 
The larger prediffusion \cite{test1} for the SHASTA visible in (a) 
is expected on account of
the smaller Courant--Friedrichs--Lewy number 0.4 as compared to 0.99 
used in the RHLLE run \cite{test1}. The sightly slower cooling of the center is
due to the larger numerical diffusion of the SHASTA. 
The sharp cusp-like structures that develop 
in the RHLLE solution in (b) and which are
associated with deflagration discontinuities in the transition from
quark--gluon to hadron matter \cite{dhrmg,mgkaj},
are therefore also broadened in the SHASTA calculation. 
(The numerical diffusion for the RHLLE is so small that it even tends to 
produce small-scale oscillations around the true solutions at the
origin and for late times in (b).)
Up to these small numerical effects, however, 
agreement is satisfactory and establishes 
confidence in that the SHASTA algorithm is able to generate approximately
correct hydrodynamical solutions also for more complicated initial conditions.

A third  test of the numerical stability of the algorithm
is shown in Fig.\ 7, comparing the experimentally observable
transverse energy flow, ${\rm d}E_\perp/{\rm d}y {\rm d}\phi$, 
and the azimuthal
correlations of the transverse energy flow, $C_{ET}(\Delta\phi)$,
in the case of ideal gas and Bag model equations of state.
An initially azimuthally symmetric Gaussian energy density profile
of radius 1 fm is evolved 100 time steps (with time step width
$\Delta t= \lambda\, \Delta x= 0.08$ fm/c).
The thermal smeared transverse energy distribution is 
computed as a function of time via eq.\ (\ref{det})
derived in the Appendix. The azimuthal transverse energy correlation
is defined by
\beq
C_{ET}(\Delta\phi)= \int_0^{2\pi}\frac{{\rm d}\phi}{2\pi}\;
 \frac{E_\perp(\phi+\Delta\phi) \; E_\perp(\phi)}{\langle E_\perp\rangle^2} 
-1\; ,
\eeq{cet}
where we abbreviated $E_\perp(\phi)\equiv 
{\rm d}E_\perp/{\rm d}y{\rm d}\phi$ and 
$\langle E_\perp \rangle \equiv \int {\rm d}\phi\, E_\perp(\phi)/2\pi$. 
For this test case, the exact solution
has $E_\perp(\phi)=\langle E_\perp\rangle$ independent of $\phi$ and 
$C_{ET}=0$. As shown in Fig.\ 7, while the initial condition 
agrees with that expectation, numerical errors develop at later times 
due to the cartesian grid that breaks the symmetry of the initial conditions.
This is especially obvious from the fact that the
azimuthal anisotropy  develops
directed along the four diagonal directions of the grid.
The numerical anisotropy is largest for the first order transition
case and reaches 10\% in the $E_\perp(\phi)$ distribution
while the numerical $E_\perp$ correlations remain below $0.1\%$. 
These results indicate the magnitude of numerical errors 
we must keep in mind in order to assess the significance of
results for more complex geometries.

\subsection{Transverse Shocks in Inhomogeneous Geometries}

In this section we demonstrate that a novel class of azimuthally asymmetric
collective flow patterns can develop from  
inhomogeneous initial conditions. These patterns are analogous to the
hadronic volcanoes proposed by T.D.\ Lee \cite{tdlee}.
However, instead of emerging from 
instabilities at the hadronization surface, these ``volcanoes''
arise from transverse directed shock waves
formed during the expansion of initial state inhomogeneities \cite{george}.

To illustrate this type of collectivity, consider
the evolution of two idealized Gaussian hot spots with radii 1 fm
and  with
centroids separated by 4 fm in the transverse $x-$direction. 
The initial transverse flow velocity is assumed to vanish
and only the Bjorken longitudinal flow velocity (\ref{ubj}) is taken
into account. The initial energy density of each Gaussian is assumed to peak
at 30 $\gfm$ to simulate the hot spots seen in Figs.\ 2--4.

In Figure 8 the evolution of the expanding hot spots
is shown for an ideal gas equation of state. 
The shaded energy density contours are shown
every 10 computational time steps, corresponding to $0.8$ fm/c between 
each frame. The expansion proceeds, as expected \cite{greinrisc},
with the formation of two (cylindrical) shells.
At 2.9 fm/c (left column, middle row), a high density wall of shocked matter
has formed at $x=0$ where the expanding shells intersect.
Subsequent evolution of that dense region leads 
to the ``eruption'' of  two back-to-back
jets of matter in the transverse $y-$direction.
These transverse shock patterns  are 
similar to the familiar ``squeeze-out''
flow pattern \cite{stocker} produced in lower energy nuclear collisions.
However, they are produced here by the collision of radially expanding shells
or bubbles of relativistic hot matter,
and in the case of multiple initial inhomogeneities, multiple
volcanoes form at azimuthal angles that depend on the particular geometry.

These transverse shocks are most clearly 
visible in the ${\rm d}E_\perp/{\rm d}y{\rm d}\phi$ 
distribution as shown in Fig.\ 9.
The initial ${\rm d}E_\perp(\tau_{th})
/{\rm d}y {\rm d} \phi \simeq 24$ GeV is of course
rotation invariant. However, at the end of the evolution
($\tau = 7.7$ fm/c), two narrow towers of directed transverse energy
emerge at $\phi=90$ and 270 degrees.
This occurs because the initial azimuthal asymmetry in coordinate space
is transferred into azimuthal asymmetry
in momentum space through evolution. Finite dissipative effects would
of course decrease the intensity and broaden the azimuthal distribution
of these volcanoes.
 
Note also that the overall magnitude of the transverse energy decreases with
time because work is performed by the fluid associated with 
longitudinal Bjorken expansion. The Gaussian hot spots expand not only in the
transverse direction but also along the beam direction.

The initial $E_\perp$--correlation function is of course zero.
By the time of freeze-out $({\cal E} \sim 0.1\; \gfm)$, however,
a strong forward and backward correlation
develops for this geometry. It is important to observe that the
magnitude of the correlation is only  30\% even though the collective
signal-to-thermal noise ratio, $s$, in ${\rm d}E_\perp(\tau_f=7.7
\, {\rm fm/c})/{\rm d}y {\rm d}\phi$ is peaked at $s \simeq 4$. 
This is a general feature of correlation functions
since the convolution introduces a dependence on the width
of the signal as well. 
If the relative width of the signal to noise is $\delta$, 
then a simple estimate of the auto-correlation is
\beq
C(0)\simeq \frac{(s-1)^2 \delta(1-\delta)}{(1+\delta (s-1))^2}
\; \; .
\eeq{c0}
We can understand qualitatively the magnitude of the correlation
at $\Delta \phi=0$ in Fig.\ 9 taking  $s \simeq 3$ and $\delta \simeq 0.2$.
The off-peak anti-correlation can also be estimated as
\beq
C(\pi/2) \simeq -\frac{(s-1)^2 \delta^2}{(1+\delta (s-1))^2}
\; \; ,
\eeq{cpi2}
which is also in qualitative accord with the computed correlation.
These correlations are numerically significant
because from Fig.\ 7 we found that
the numerical errors lead only to a $0.01\%$ correlation in 
the ideal gas case. The slight forward-backward asymmetry
in the correlation function is only due to 
our histogram binning procedure.

In Figure 10 we show the evolution of the same initial condition assuming 
the Bag model equation of state, eq.\ (\ref{bag}).
The expansion in this case is qualitatively different from
the ideal gas case. The expanding shells or bubble walls
are much thinner as is also the high density
intersection region. This is even more clearly seen in Fig.\ 11
that shows the energy density as a function of time along the $x-$axis for
both equations of state. The transverse velocity profile along
that slice is also shown. In the case of a first order transition,
a sharp cusp is produced at $x\simeq -5,0,5$ fm, which correspond
to points where matter cools to the 
critical mixed phase transition point, $\epsilon_Q$.
In contrast, the bubbles and transverse shock zones in the ideal gas case 
remain comparable to the width of the initial hot spots. Those structures
are much thinner in the phase transition case. The reason is that
mixed phase matter does not resist compression due to the vanishing
of the speed of sound and therefore fills less space than an ideal
gas with finite compressibility (see also Fig.\ 6).

Also, the evolution in the case of a first order transition 
is slower by about a factor of two relative to Fig.\ 8. This is due to 
the stall of the expansion in the mixed phase because
of the vanishing velocity of sound in those regions 
\cite{risch_pipi,dhrmg}.
In the ideal gas case, we have $c_s^2=1/3$ throughout the expansion.
With $c_s=0$ in the region with $\epsilon_H <\epsilon<
\epsilon_Q$, sharp cusps are formed that move outwards only
slowly.
As can be seen from the velocity profiles the flow velocity has a discontinuity
through the cusp, typical of deflagration phenomena 
\cite{risch_pipi,dhrmg,mgkaj}.

The hydrodynamic stability analysis of bubble formation is complex
and in the cosmological electroweak 
context is still subject to controversy \cite{kajmec,kamion}.
In the QCD case even less is known, 
though at least in \cite{kajmec} marginal stability was found to be possible
within the uncertainties of the relevant scale.
In our case, bubble formation does not result from supercooling 
but rather from the dynamical expansion of initial state
inhomogeneities. Whether these hydrodynamic structures are stable is left
here as an open question,
especially since the thickness of the bubble walls is of hadronic dimensions.
As we show in the next section the stalled expansion and the formation
of thin shells of expanding mixed phase matter is the typical pattern
we find also for the more complex inhomogeneous, turbulent mini-jet initial 
conditions.

Returning to Fig.\ 9, we see that the consequence of stalled expansion
is a considerable reduction of the transverse shock intensity
as measured by ${\rm d}E_\perp/{\rm d}y {\rm d}\phi$, 
relative to the ideal gas case.
The signal to noise is reduced to $s\simeq 1.5$ and the relative width is
also reduced to $\delta \simeq 0.1$. With these reductions, the $E_\perp$
correlation is seen to be reduced by an order of magnitude in accord
with eqs.\ (\ref{c0},\ref{cpi2}). However, the few--percent correlation
is still numerically significant compared to the $0.1\%$ numerical
errors deduced from Fig.\ 7. 

\subsection{Evolution of Turbulent Mini-Jet Initial Conditions}

In Figures 12, 13 we compare the evolution of a typical mini-jet
initial condition. The initial $T^{\mu\nu}$
required for the hydrodynamic evolution is taken from
eq.\ (\ref{xebj}) with $T^{00} \equiv {\cal E}, \, T^{0i} \equiv M^i$. 
We note that the HIJING
initial conditions are not of the ideal fluid form. 
By only taking the left hand
side of (\ref{xebj}) to fix all other components
of the fluid energy--momentum tensor, we convert the HIJING initial conditions
into an ideal fluid form through the assumption
of thermalization at $\tau_{th}$. Thermalization has the effect of {\em
reducing\/} the initial transverse energy somewhat from the HIJING input,
because some of the transverse energy is converted into longitudinal thermal
motion.

The time steps between frames in the case of a first order transition
(Fig.\ 13) 
are taken to be twice as long (1.6 fm/c) as in the ideal gas case (Fig.\ 12) 
to take into account the inherently slower and stalled expansion
in the first order transition case. 
For this event, several prominent hot spot regions are seen to expand 
in a manner analogous to the previous Gaussian examples. However, in 
this case the hot spots also start with initial transverse collective flow
velocities determined from the mini-jet fluctuations. 

The background fluid produced via soft beam-jet fragmentation
in this example is assumed to have the smooth, azimuthally symmetric profile
\beq
{\cal E}^{\rm soft}(\tau_{th},\vxp)= \frac{{\rm d}E^{\rm soft}_\perp}{
{\rm d}y}\,
\frac{1}{\tau_{th} \pi R^2}\, \frac{3}{2} (1-r_\perp^2/R^2)^{1/2} \; \; ,
\eeq{epssoft}
with ${\bf M}_\perp^{soft}=0$ as well.
Unlike in Figs.\ 2--4 where we included only about one
half of the soft transverse energy on account of the formation time
estimates, in this case we have taken the full
${\rm d}E^{\rm soft}_\perp/{\rm d}y=0.5$ TeV estimated from HIJING. 
The soft component adds in this case (i.e.\ $p_0=2$ GeV/c) 
a larger smooth background of depth 
${\cal E}^{\rm soft}(\tau_{th},0) \simeq 10\; \gfm$ to the central region.
As emphasized  before, the magnitude of that soft background component
is uncertain to about a factor two,
and we take the above value to be on the conservative side.

In Figure 12 one can see two main expanding bubbles emerging from the 
two hottest spots. Their evolution is more irregular than in Fig.\ 8
because of the inhomogeneous background in which they propagate.
The solid curve shows the freeze-out hadronization contour
with $\epsilon=\epsilon_H=0.1\;\gfm$. In this ideal gas case,
this hadronization surface shrinks monotonically but does not correspond
to any obvious ${\cal E}$ contour because of the underlying chaotic 
collective flow velocity field.

In Figure 13, in addition to slowing down
of the expansion, the evolution with the first order transition
leads to the production of much thinner and irregular
bubble fragments of mixed
phase matter reminiscent of the thin shells in Fig.\ 10.
It is the multiple inhomogeneities and shear velocity fields that
make these structures much more irregular in this case.
The hadronization surface ($\epsilon(\tau,\vxp)=\epsilon_H$) also seems to
acquire a foam-like structure\footnote{Bubble formation 
is a natural characteristic of a mixed phase, the above foam, however,
is induced by the initial inhomogeneities and the subsequent collective
motion.}.
The hadronization surface tends in this case of a first
order transition to coincide
closer with the plotted ${\cal E}(\tau,\vxp)=T^{00}(\tau,\vxp)$ contours 
because the velocity of the bubble walls are smaller
than in the ideal gas case while the ejected cooler hadronic matter
tends to have higher flow velocity (see \cite{mgkaj}).

In Figure 14, a cut along the $x-$axis 
provides a close-up of the complex nature
of the sharp structures of mixed phase matter that emerge
as well as of the chaotic transverse velocity fields in between them.
It also shows the energy density scales corresponding
to the shaded contours in Figs.\ 12 and 13. In this view, the
sharp initial inhomogeneities due to mini-jets superimposed
on the smooth beam-jet background and the initial turbulent velocity field
are particularly clearly revealed.

The evolution of the transverse energy in this event is shown in Fig.\ 15.
Unlike in the static examples discussed before,
the initial ${\rm d}E_\perp/{\rm d}y {\rm d}\phi$ 
is not azimuthally symmetric in this
case because of the initial turbulent velocity field. Of course,
the azimuthal angles of the bumps and valleys vary from event to event. 
That is why correlation functions must be studied experimentally. However,
the evolution of ${\rm d}E_\perp/{\rm d}y{\rm d}\phi$ 
in this event reveals the general tendency of inhomogeneous
initial conditions  to evolve into multiple azimuthally directed flow
structures.
Comparing the ideal gas and first order transition cases,
we see again that the former leads to a lower average final transverse energy
than the latter due to extra work done by the ideal gas
upon longitudinal expansion.

In Figure 16 we show ${\rm d}E_\perp/{\rm d}y{\rm d}\phi$ 
averaged over 50 events 
and $C_{ET}(\Delta\phi)$ for such turbulent initial conditions.
The event-averaged ${\rm d}E_\perp/{\rm d}y{\rm d}\phi$ 
remains approximately azimuthally symmetric as required. Initially, 
there is a small ($<1\%)$ azimuthal 
auto-correlation that is induced when the HIJING 
parton data are coarse-grained
into 1 fm$^3$ fluid cells. The initial state correlation
also includes the small fluctuating dipole contribution
arising from the fact that $\sum \vpa\ne 0$ for a finite number
of mini jets in the central rapidity slice.
At later times, however, a $3\%$ auto-correlation
develops in the ideal fluid case and approximately $2\%$
in the first order transition case. 
The magnitude of the correlations for both equations of state
is evidently small.

In Figure 17, the dependence of the induced $E_\perp$ correlations
on the soft background as well as on the mini-jet scale parameter
$p_0$ is shown. The HIJING default case from
Fig.\ 16 corresponds to the solid curves.
In the ideal gas case (left panels), the auto-correlation reaches
$6\%$ if the background is reduced by a factor of two (dashed curve)
to ${\rm d}E_\perp^{soft}/{\rm d}y
=250$ GeV as in Figs.\ 2--4. On the other hand, with $p_0=2$ GeV/c fixed
but the soft background increased by a factor of two (dotted curve)
to ${\rm d}E_\perp^{soft}/{\rm d}y
=1$ TeV, the final auto-correlation is reduced by around a factor of two
relative to the default case.
Finally, if $p_0$ is decreased to 1 GeV/c but
${\rm d}E_\perp^{soft}/{\rm d}y
=0.5$ TeV is kept fixed (dash--dotted curve),
the correlation hardly changes relative to the corresponding
$p_0=2$ GeV/c case (solid).  The wiggle in this last case
is due to the more limited statistics (20 events) available for this average.
We conclude that in the ideal gas case, the collective azimuthal anisotropies
are approximately linearly dependent on the level of
the soft beam-jet component.

The initial state correlation function, however, also depends on the
$p_0$ and soft background scales. In the lower panels,
the ratio of the final correlation function to the initial one is shown
in the restricted $\Delta \phi< \pi/2$ interval that avoids the
artificial pole created at the point where the
initial correlation function crosses zero.
It is interesting to note that this ratio for the three $p_0=2$ GeV/c
curves is practically independent of the background level.
On the other hand, the $p_0=1$ GeV/c ratio is significantly larger.
Thus, while the absolute magnitude of the correlation function
depends roughly linearly on the soft background level, the dynamical
enhancement of the initial state correlations in the ideal gas case
peaks near 4--5 at $\Delta \phi=0$ approximately independent 
of that background. 

While the absolute collective signature as measured by the 
$E_\perp$--correlation
function is small, it is quite significant compared to the initial state
correlations (Fig.\ 16) and the numerical accuracy (Fig.\ 7).
The transverse energy correlation function
is of course only one way
to search for azimuthally asymmetric collective flow phenomena.
The power spectrum, $E_\perp(m)
=\langle \int {\rm d} \phi\; e^{-i m \phi}\, E_\perp(\phi)\rangle$, 
wavelet analysis, and factorial moment fluctuation analysis may provide more
sensitive probes of the induced collectivity that is apparent
in the ratio curves in Fig.\ 17.

In the case of the first order phase transition,
the $E_T$ correlations are significantly suppressed relative to the ideal gas
case and appear to be much more insensitive to the background level.
Comparing the solid and dotted curves 
indicates, however, a stronger dependence
on the $p_0$ mini-jet scale.
For $p_0=1$ GeV/c and ${\rm d}E_\perp^{soft}/{\rm d}y
=1$ TeV, the azimuthal anisotropies fall below $1\%$.
On the other hand, in the ratio of final to initial 
correlation functions, the largest enhancement occurs for the $p_0=1$ 
GeV/c case.
The ratios also show a qualitative shoulder feature in the first order 
transition case for which we have not found a simple explanation. The
level of collectivity is, however, probably too small to see such structures.

The suppression of collective flow phenomena in the case of a first order
transition is due to the vanishing of the speed of sound over
a large interval of energy densities, $\epsilon_H<\epsilon<\epsilon_Q$.
This is also manifest in the smaller reduction of the initial
${\rm d}E_\perp/{\rm d}y$ due to longitudinal expansion
relative to the expansion with an ideal gas equation of state.
The above results can be understood as a consequence
of a rather general feature of evolution with any equation
of state that possesses a dip of  $c_s^2={\rm d}p/{\rm d}\epsilon$ 
in a finite range of energy densities. In the case of a first order 
transition, $c_s^2=0$ 
in the mixed phase, and pressure gradients driving collective flow phenomena
are strongly suppressed. 
As emphasized in \cite{risch_pipi}, even a continuous
cross-over transition may feature  such a minimum if the
cross-over temperature region is not too broad. For realistic
$\Delta T/T_c \simeq 0.1$, the softening of the equation of state
is sufficiently strong that the time-delay signature discussed in the next 
section should still be observable.
The same physics of softening is expected to lead
to a suppression of directed transverse flow phenomena at much lower
AGS energies \cite{risch_flow}
and also to the suppression \cite{vanhove} of
${\rm d}\langle p_\perp\rangle/{\rm d}({\rm d}N_\pi/{\rm d}y)$ 
in the mixed phase region. In the turbulent-glue scenario,
the suppression of pressure gradients
in the cross-over region of the equation of state 
has the observable consequence of reducing the azimuthally 
asymmetric collective phenomena. 

It is indeed curious that many ``barometric''
signatures of QGP formation involve
the {\em suppression\/} of collective behaviour that would otherwise
naturally arise in ideal gas systems. This makes the search for
signatures of the QGP more difficult because ordinary dissipative effects
due to viscocity, thermal and colour conductivity etc.\ work in the same
direction. It is only through the careful systematic studies
of all observables as a function of beam energy and nuclear size
that one can hope to unravel interesting threshold-type 
behaviour caused by the passage through a mixed-phase region
from the suppression of collective phenomena due to 
less interesting dissipative dynamics.
 
\section{Robustness of the Time-Delay Signature}

Since it is the reduction of collective observables in the evolution of a QGP
that signals rapid cross-over regions in the equation of state,
it is important to find as many correlated signatures
as possible in the search for that phase of matter. 
As repeatedly stressed above,
one of the generic consequences of hydrodynamics
with an equation of state that has a soft region (reduction of $c_s^2$)
is time delay. Meson interferometry has been proposed
\cite{pratt,bertsch} as the main experimental tool to search
for such time-delay signatures.
As shown recently in \cite{risch_pipi}, that signature of stalled 
dynamics is fortunately robust to an increase in the width
of the cross-over region. In this section,
we want to demonstrate in more detail the robust character of that
observable even to the much more unfavourable
turbulent initial conditions discussed above.

In Figure 18, the evolution of the {\em mean\/} energy
density is shown, averaged over the inner $r_\perp<3$ fm core of the plasma.
The average is again over 50 events for each equation of state.
At $\tau=\tau_{th}=0.5$ fm/c, the mean central energy density
is approximately $16\; \gfm$ for the turbulent ensemble with $p_0=2$ GeV/c and
${\rm d}E_\perp^{soft}/{\rm d}y =0.5$ TeV. The solid curve, for the case of a
first order transition and 3+1--dimensional expansion, should be compared to
the thick dashed curve for the case of an ideal gas equation of state.
In addition, the light dashed and dash--dotted curves are shown
for comparison. They correspond to 
ideal one--dimensional Bjorken expansion
with transverse expansion neglected.
The $\tau^{-1}$ curve represents pure longitudinal
expansion without work, $p=0$,
while the $\tau^{-4/3}$ curve corresponds to the ideal gas case.
The 3+1--dimensional ideal gas evolution starts to deviate from the 
one--dimensional case after a short
time $\sim 2$ fm/c due to rapid radial expansion. In the first order transition
case, the mean energy density follows the ideal one--dimensional Bjorken 
curve up to $\sim 6$ fm/c because the transverse expansion is stalled.
The freeze-out occurs near $\epsilon\simeq \epsilon_H\simeq 0.1\;\gfm$.
It is clear from Fig.\ 18 
that the freeze-out time is approximately twice as long in
the case of a phase transition to the QGP. 

Especially important for the
$R_{\rm out}/R_{\rm side}$ signature \cite{pratt,bertsch,risch_pipi} of the QGP
is that the transverse coordinate distribution in the first order transition
case remains more compact even though it takes a longer time for the system
to freeze-out. This can be seen clearly in Figs.\ 12 and 13
which, together with Fig.\ 18, confirm that 
the space--time geometry of freeze-out
remains so different in the two cases. 

Figure 19 emphasizes the strongly inhomogeneous
character of expansion in the turbulent-glue scenario. The thick solid
and dashed curves are the same as in Fig.\ 18. The two thin curves
show the magnitude of the rms fluctuations of the energy density
in the central core for the first order transition case.
The initial fluctuations are already large. However, those fluctuations
grow rapidly as the system passes through
the mixed phase and the foam structure
in Fig.\ 13 develops. This shows that freeze-out is not a fixed-time event,
but is spread over the entire evolution.
The most essential aspect for the time-delay signal is 
that the freeze-out time duration is large relative to the transverse 
coordinate dispersion \cite{pratt,bertsch,risch_pipi}.

The actual computation of the pion interference pattern from the turbulent
evolution is beyond the scope of the present study. A substantial 
generalization of the already computationally demanding methods used in 
Ref.\ \cite{risch_pipi} will be required. However, based on that work,
we expect the general interference pattern to reflect well the underlying,
time-delayed freeze-out geometry. Therefore,
the time-delay signature survives even if the initial conditions
are as inhomogeneous as in the turbulent-glue scenario.

The detailed interference pattern can be expected, on the other hand,
to differ from the patterns produced from the evolution of
homogeneous initial conditions due to the overlapping
bubble wall geometries found in Fig.\ 13. The Fourier transform of
such hadronic foam is bound to contain extra structure since
multiple distance scales (shell thickness and radii, and relative separations)
enter the coordinate space distribution.
It would be interesting in the future to explore such novel interference
patterns that would be specific to inhomogeneous geometries.

\section{Enhanced Radiance of Hot Spots}

A specific consequence of hot-spot formation is enhanced
radiance of hard probes. For instance, the (invariant) rate for emitting
photons with momentum ${\bf k}$ from a fluid element consisting of
quarks and gluons
with temperature $T$ and 4--velocity $u^{\mu}$ is given by \cite{kapusta}
\beq
\omega\, \frac{{\rm d} R^{\gamma}}{{\rm d}^3 {\bf k}} = 
\frac{5 \alpha \alpha_S}{18 \pi^2}\;T^2
e^{-k \cdot u/T} \ln \left( \frac{2.912\, k \cdot u}{g^2 T} + 1 \right)\;\;.
\eeq{phot}
It was moreover shown in \cite{kapusta}
that this rate is approximately the same if the fluid element
consists of hadrons instead of QGP.
To estimate photon radiation in the turbulent-glue scenario as
compared to that from a homogeneous Bjorken cylinder,
we integrate numerically the rate 
(\ref{phot}) over the photon emission 
angle, over (proper) time and transverse coordinates, and evaluate it at
$y=\eta = 0$ ($\eta= \ln[(t+z)/(t-z)]/2$ being the space--time rapidity).

The final differential photon spectrum
${\rm d}N^{\gamma}/{\rm d}\eta {\rm d}y k_\perp {\rm d}k_\perp |_{y=\eta=0}$ 
is shown in Fig.\ 20
averaged over 10 HIJING events (thick lines) and for a homogeneous 
Bjorken cylinder (thin lines) of radius $R=4$ fm with the same initial 
(average) energy as the HIJING events.  We note that
even a single event
produces a quite similar spectrum as the 10--event average.
As one expects, the higher
temperatures of hot spots in the turbulent-glue scenario lead to an 
enhancement of the exponential tail of the photon spectrum. 

Comparing the yield in the case of a  phase transition (dashed lines) to that
in the case of an ideal gas equation of state (solid lines), one observes 
that the longer lifetime of the system in the case of a phase transition
leads to enhanced radiation of photons with small as well as
with large momenta in the case of the turbulent-glue scenario, while
for the homogeneous Bjorken cylinder, only radiation of
photons with small momenta is significantly enhanced. 
The higher fluid velocities
in the case of the first order transition scenario as well as
reheating in the shocks created in the expansion of hot spots 
seem to be responsible for the stronger population of the 
high-momentum tail of the spectrum.
Correlations of direct gammas with azimuthally directed transverse shocks
should also be looked for.
Other probes such as strangeness enhancement may also be correlated
with transverse shocks.

\section{Summary}

In this paper we studied  several possible  consequences of initial state
inhomogeneities and turbulence in quark--gluon plasmas arising from
multiple mini-jet production in ultrarelativistic nuclear collisions.
The HIJING model was used to generate the initial mini-jet configurations
and to provide an estimate of the soft beam-jet background.
The ensemble of initial conditions was found to exhibit a wide
spectrum of fluctuations not only of the initial energy density
distribution but also of the initial
transverse velocity field. The fluctuations are large 
if the equilibration time of the plasma is short, $\sim 0.5$ fm/c,
as suggested by recent estimates of radiation energy loss.
We refer to this type of initial conditions as the
turbulent-glue scenario to contrast it with the more conventional
hot-glue scenario which assumes cylindrical symmetry,
homogeneity, and quiescence of the initial plasma.

We assumed the validity of non-dissipative
hydrodynamics in order to assess the most optimistic observable
consequences of such unfavourable initial conditions.
We studied three observables that could serve as diagnostic
tools of such plasmas. 

First, we showed that new types of azimuthally asymmetric
collective flow patterns (volcanoes) could arise in inhomogeneous
geometries. At the event-by-event level they could be observed
by looking for spikes
in the transverse energy distribution, ${\rm d}E_\perp/{\rm d}y{\rm d}\phi$.
However, the collective effects are difficult to identify
because of large uncorrelated fluctuations associated with the
turbulent nature of the initial conditions.
Those uncorrelated fluctuations can be averaged out
by studying instead the azimuthal 
correlation function of the transverse energy.
The remaining dynamical correlations are small
but sensitive to the plasma equation of state. In general, 
evolution with an equation of state featuring a minimum of the speed of sound
in the energy density range of interest reduces all collective 
flow phenomena relative to the ideal gas case. For example, the overall
reduction of the initial transverse energy due to work associated with
longitudinal expansion is maximal for the ideal gas case.

Second, we discussed the time-delay signature of a QGP transition.
We found that, as in the conventional hot-glue scenario,
the evolution is stalled if there is a soft region of the equation of state.
It is remarkable that this phenomenon survives not only if 
the QGP equation of state only features a smooth cross-over instead
of a first order transition \cite{risch_pipi,dhrmg},
but also if the 
ensemble of initial conditions is as complex as in the turbulent 
glue-scenario. Meson interferometry 
\cite{pratt,bertsch,risch_pipi} appears  therefore
to be one of the most robust diagnostic tools in the search for
the QGP transition.

Finally, we considered briefly the effect of hot spots on hard probes.
The thermal contribution to probes such as direct photons is obviously
exponentially sensitive to the local temperature. We showed that
the turbulent-glue scenario can enhance by more than an order
of magnitude the high--$k_\perp$ tails for both equations of state.
Other hard probes and even softer ones such as strangeness production
can be expected to be correlated and enhanced due to hot spots
and collective transverse shock phenomena. However, to identify
any enhanced yields with hot-spot formation
will require subtraction of 
other pre-equilibrium contributions to those yields.

The above scenario represents in many ways the most
optimistic point of view for signatures of QGP formation.
We neglected all dissipative effects that tend
to dampen collective behaviour of the system and decrease the 
freeze-out time. The thin expansion shell structures found may be diffused
considerably by such effects. 
Chemical equilibrium may not be maintained through
the transition point. In general, chemical rate equations
would have to supplement the hydrodynamic equations.
Also, we assumed Bjorken boundary conditions and hence
neglected fluctuations along the longitudinal
direction. In the general 3+1--dimensional case,
fluctuations will produce hot-spot droplets 
instead of the cylindrical structures studied here.
That extra dimension decreases the overlap phase space
of expanding inhomogeneities and therefore reduces
the pattern of collective transverse shocks. 
We have also assumed a most conservative homogeneous beam-jet
background underneath the turbulent mini-jet plasma.
In confronting data eventually, 
these and other  complications will have to be considered in more
detail in future studies.\\[2ex]

\noindent 
{\bf Acknowledgments}
\\ ~~ \\ 
We thank M.\ Asakawa, F.\ Cooper, C.\ Greiner, J.\ Harris, D.\ Heumann,
B.\ K\"ampfer, J.\ Kapusta, K.\ Kinder--Geiger, T.D.\ Lee, 
B.\ M\"uller, P.V.\ Ruuskanen,
E.\ Shuryak, H.\ St\"ocker, and X.N.\ Wang for stimulating 
discussions during the course of this work. 
\\ ~~ \\
\appendix
\section*{Appendix: The Transverse Energy Flow}

The local thermal phase space density for a
mixture of massless fermions and bosons is
\beq
f(t,\vx,\vp)=\frac{{\rm d}^6N}{{\rm d}^3\vx\, {\rm d}^3\vp} 
=\frac{1}{(2\pi)^3}\left(
\frac{g_B}{e^{p \cdot u/T}-1} +
\frac{g_F}{e^{p \cdot u/T}+1}\right)  
\equiv h\left(\frac{p \cdot u}{T}\right) \; \; ,
\eeq{feq}
where $T(x)$ and $u^\mu(x)$ are the local
temperature and fluid 4--velocity, and $g_B(g_F)$ are the number
of active massless Boson (Fermion) helicity states.
For example, in the mixed phase 
$g_B=16 r_Q + 3(1-r_Q)$ if the fraction  of matter in the plasma
phase is $r_Q$.
The fluid energy--momentum tensor is related to $f$ via
\beq
T^{\mu\nu}(x)=\int \frac{{\rm d}^3\vp}{p^0}\; p^\mu p^\nu 
h\left(\frac{p \cdot u}{T}\right) \; \; .
\eeq{eps}
For longitudinal boost-invariant (Bjorken) boundary conditions,
$T=T(\tau,\vxp)$, depends only on the transverse coordinates
$\vxp$ and the proper time
$\tau=\sqrt{t^2-z^2}$. The longitudinal fluid velocity
is determined by the space--time rapidity variable
$\eta=\ln[(t+z)/(t-z)]/2$, via  $v^z=u^z/u^0= \tanh \eta$.
If the transverse flow velocity at $z=0$ ($\eta=v^z=0$) 
is specified as ${\bf v}_\perp= \tanh y_\perp (\cos \phi_\perp,
\sin \phi_\perp)$ in terms of the local transverse flow rapidity
$y_\perp(\tau,\vxp)$ and its azimuthal direction by the
angle $\phi_\perp(\tau,\vxp)$, then the fluid
4--velocity is given by
\beqar
u^\mu(\tau,\vxp,\eta)&=& (\cosh \eta \cosh y_\perp , 
\sinh y_\perp \cos \phi_\perp,
\sinh y_\perp \sin \phi_\perp,\sinh \eta \cosh y_\perp)\nonumber \\
&=& \gamma_\perp (\cosh \eta , 
v_\perp \cos \phi_\perp,
v_\perp \sin \phi_\perp,\sinh \eta) \;\; ,
\eeqar{umu}
where $\gamma_\perp \equiv \cosh y_\perp$ and $v_\perp\equiv\tanh y_\perp$.
Note that if the  transverse flow velocity vanishes,
$y_\perp=0$, then we recover the familiar Bjorken flow velocity
field $u^\mu=x^\mu/\tau$.

It is convenient to express $p^\mu$ in terms
of longitudinal rapidity and cylindrical transverse momentum variables,
such that $p^\mu=p_\perp(\cosh y, \cos \phi, \sin\phi,\sinh y)$.
Therefore, the argument of the thermal distribution $h$ above reduces to
\beq 
p \cdot u/T =p_\perp \gamma_\perp (\cosh (y-\eta)
- v_\perp \cos (\phi-\phi_\perp))/T 
\eeq{put}
It is linearly dependent on $p_\perp$ because we assume that all components
of the fluid are massless. 

The energy density at $z=0$, i.e. $\eta=0$, is then given by
${\cal E}=T^{00}$ via 
\beqar
{\cal E}(\tau,\vxp,0)=\left(\frac{T}{\gamma_\perp}\right)^4 
\int_{-\infty}^\infty {\rm d}y\, \cosh^2 y
\int_0^{2\pi} \frac{{\rm d}\phi}{ (\cosh y
-v_\perp \cos\phi)^{4}} \int_0^\infty {\rm d}x x^3 h(x) \;\; .
\eeqar{ep}
Note that 
\beq
\int_0^\infty {\rm d}x x^3 h(x)= \frac{1}{4\pi}(g_B+7g_F/8)
\frac{\pi^2}{30}\;\; .
\eeq{int1}
When this last factor is combined with
$T^4$ we obtain the local {\em proper\/} 
energy density $\epsilon(\tau,\vxp)$ $/(4\pi)$.
In the limit where $v_\perp(\tau,\vxp)=0$, the remaining two integrals
then yield a factor $4\pi$ and thus eq.\ (\ref{ep}) reduces to the
expected 
\beq
\lim_{v_\perp\rightarrow 0} {\cal E}(\tau,\vxp,0)= 
(g_B+7g_F/8)\; \frac{\pi^2}{30}\; T^4(\tau,\vxp)
\equiv \epsilon(\tau,\vxp) \;\; .
\eeq{eps00}
When the local transverse flow velocity does not vanish,
the last two integrals are still analytic
and implement the correct  Lorentz transformation
of the energy density in the ideal fluid limit
with  
\beq
{\cal E}(\tau,\vxp,0)=\frac{1}{3}\,(4\gamma_\perp^2 -1)\, \epsilon(\tau,\vxp)
\;\; .
\eeq{eps1}
(Recall that for an ideal fluid $T^{\mu\nu}=
(\epsilon+p)u^\mu u^\nu-pg^{\mu\nu}$.)

If the fluid is frozen out at a fixed proper time, $\tau_f$,
then the invariant total particle distribution
is obtained by integrating over all fluid cells
via the Cooper--Frye formula
\beqar
p\, \frac{{\rm d}N}{{\rm d}^3{\bf p}}&=&\frac{{\rm d}N}{{\rm d}y
\, {\rm d}^2{\bf p}_\perp}
=\int_{\Sigma_f} {\rm d} \sigma^\mu\, p_\mu \; f(\tau_f,\vx,\vp)\nonumber \\
&=& \tau_f p_\perp\; \int {\rm d}^2\vxp \int {\rm d}\eta\, \cosh(\eta-y)\,
h\left(\frac{p \cdot u}{T}\right) \; \; ,
\eeqar{invn}
where we used the 3--volume element 
\beq
{\rm d}\sigma^\mu =\epsilon^{\mu\alpha\beta\gamma}\,
\frac{\partial x_\alpha}{\partial\eta}
\frac{\partial x_\beta}{\partial x_\perp}
\frac{\partial x_\gamma}{\partial y_\perp}\, {\rm d}\eta \, {\rm d}^2\vxp
=(\cosh\eta,0,0,\sinh\eta) \; \tau_f\, {\rm d}\eta \, {\rm d}^2\vxp
\eeq{dsig}
given $x^\mu(\tau,\vxp,\eta)=(\tau\cosh\eta,\vxp,\tau\sinh\eta)$.
Note that ${\rm d}N/{\rm d}y{\rm d}^2{\bf p}_\perp$ is of course
independent of $y$ given our boundary conditions, but if $y_\perp
(\tau,\vxp)\ne 0$ it could be highly azimuthally asymmetric.

Finally, the transverse energy distribution at proper time
$\tau_f$ and $y=0$ is given by
\beqar
\frac{{\rm d}E_\perp}{{\rm d}y {\rm d}\phi} &=& 
\int_0^\infty {\rm d}p_\perp\, p_\perp^2 \frac{{\rm d}N}{{\rm d}y
{\rm d}^2 {\bf p}_\perp}
=\tau_f \int {\rm d}^2\vxp
\int {\rm d}\eta\;\cosh \eta \int {\rm d}p_\perp\;  p_\perp^3 
h \left(\frac{p \cdot u}{T}\right) \nonumber \\
&=& \tau_f \int {\rm d}^2\vxp\;\frac{\epsilon(\tau_f,\vxp)}{4\pi
\gamma_\perp^4}
\int {\rm d}\eta\frac{\cosh\eta}{(\cosh\eta-v_\perp\cos(\phi-\phi_\perp))^4} 
 \nonumber \\
&=& \int {\rm d}^2\vxp \,\tau_f {\cal E}(\tau_f,\vxp)\; \Delta(\phi-
\phi_\perp(\tau_f,\vxp),v_\perp(\tau_f,\vxp))\;\; ,
\eeqar{det}
where the smeared-out azimuthal angle delta function is given by 
\beqar
\Delta(\phi,v)&=&
\frac{3}{4\pi \gamma^4(4\gamma^2-1)}
\int_{-\infty}^\infty {\rm d}\eta\; \frac{\cosh\eta}{
(\cosh\eta-v\,\cos\phi)^4} 
\nonumber \\
&=&
\frac{1}{4\pi}\frac{(1-v^2)^3}{3+v^2}
\left\{ \frac{w(2w^2+13)}{(1-w^2)^3}+\frac{6+24 w^2}{(1-w^2)^{7/2}}
\arctan \sqrt{\frac{1+w}{1-w}}\right\} \;\; ,
\eeqar{del}
where $w=v\cos\phi$. In the $v=0$ limit, $\Delta(\phi,0)=1/8$ which is
$\pi/4$ times smaller than the naive $1/(2\pi)$ uniform distribution
expected in the non-thermal Bjorken limit.
In the $v\rightarrow 1$ limit, we can use the integral \cite{GR}
\beq
\int_0^{2\pi}\frac{{\rm d}\phi}{ (a
-b \cos\phi)^{4}} =\pi\frac{a(2a^2 +3 b^2)}{(a^2-b^2)^{7/2}}
\eeq{int2}
to show that $\int {\rm d}\phi\, \Delta(\phi,v\rightarrow 1)=1$, i.e.,
\beq
\lim_{v\rightarrow 1} \Delta(\phi,v) =\delta(\phi)
\; \; .
\eeq{deldel}

\newpage

\noindent 
{\bf Figure Captions:}
\\ ~~ \\
{\bf Fig.\ 1:} HIJING1.3 model \cite{wang} predictions 
for central ($b=0)$ $Au+Au$
collisions at $\sqrt{s}=200$ AGeV are shown
for the mini-jet cut-off scale $p_0=1$ GeV/c (solid) and 2 GeV/c (dashed). 
The rapidity density of 
mini-jet gluons including initial and final state  radiation 
is shown in part (a). Shadowing and jet quenching are not
included in these calculations.
The gluon transverse energy per unit rapidity
is shown in part (b).
In part (c) the initial transverse momentum distribution of the mini-jet gluons
at $y=0$ is shown. The final 
hadronic transverse energy distribution including
both mini-jet and beam-jet fragmentation is shown in (d).
\\ ~~ \\
{\bf Fig.\ 2:} (a) Hot spots in $Au+Au\; (\sqrt{s}=200$ AGeV, $b=0)$
at the thermalization
time $t=\tau_{th}=0.5$ fm/c are seen as spikes in the
local energy density, ${\cal E}(\tau_{th},\vxp)$, as a function
of the transverse coordinate, $\vxp$, in the $z=0$ plane.
The transverse and longitudinal resolution
scales are fixed to be $\Delta r_\perp=1$ fm and $\Delta y=1$. 
The mini-jet scale here is $p_0=2 $ GeV/c.
(b) The chaotic initial transverse momentum field, 
${\bf M}_\perp(\tau_{th},\vxp)$, is represented here by arrows.
(c) The spectrum of proper energy density fluctuations at $\tau_{th}$
in the central region with $r_\perp<4$ fm is shown.
The histogram is an average of 200 HIJING events 
and includes a soft gluon component due to beam-jet fragmentation.
(d) The spectrum of effective gluon temperatures in the central region
corresponding to (c) is shown.
\\ ~~ \\
{\bf Fig.\ 3 :} A surface plot of the local energy density ${\cal E}(\tau_{th},
\vxp)$, averaged over 200 events,
shows the smooth, azimuthally symmetric form usually assumed in the hot-glue
scenario. The shaded contour on the top part is a 2--dimensional 
projection of that 
surface plot. In parts (b)--(d), three separate HIJING events show that
the fluctuations in Fig.\ 2a are typical. The 
shaded contours illustrate  the complex nature
of the azimuthally asymmetric
inhomogeneities produced by the mini-jet mechanism.
\\ ~~ \\
{\bf Fig.\ 4 :} The gluon number densities corresponding to the
same events as in Fig.\ 3 show that the hot spots in Fig.\ 3 are not caused
by isolated jets but occur as accidental coincidence of several gluons
in the same region.
\\ ~~ \\
{\bf Fig.\ 5:} Evolution of a cylindrically symmetric Gaussian hotspot of
radius 2 fm and peak energy density ${\cal E} = 30 \; \gfm$. Shown
are (calculational frame) energy density profiles for (a) the initial
situation at time $t_0=0.5$ fm/c, (b) at time $t=14.9$ fm/c
after evolving with the standard antidiffusion fluxes and with 
the ideal gas equation of state,
(d) at the same time after evolving with antidiffusion fluxes
reduced to 70\% of the standard value, and (c) at the same time after
an evolution with the equation of state with a phase transition and reduced
antidiffusion. (b)--(d) show the profiles only for positive $x, y$, the other
quadrants are symmetric.
\\ ~~ \\
{\bf Fig.\ 6:} (a) Energy density profiles along the $x-$axis for 
times $t= \tau_0 + 1.6\, n$ fm/c, $\tau_0 = 0.5$ fm/c, 
$n=0,1,...,9$, calculated with the ideal gas equation of state.
Full lines are results from the RHLLE, dots are cell values for the
SHASTA run. (b) Energy density profiles for times $t=\tau_0
+3.2\, n$ fm/c, $n=0,1,..,4$ and $t= 14.9$ fm/c, calculated with
the equation of state with a phase transition.
\\ ~~ \\
{\bf Fig.\ 7 :} The evolution of numerical errors due to the
cartesian grid is
shown for the expansion of a single Gaussian initial condition.
The transverse energy fluctuations are initially azimuthally symmetric 
but develop
10 \% systematic asymmetries in the case of a first order transition
(upper right panel).
The maximal $E_\perp$ correlation induced by the grid is, however, only on the
order of 0.1\% (lower right panel). In the ideal gas case, errors are about
an order of magnitude smaller (left panels) at the same time.
\\ ~~ \\
{\bf Fig.\ 8 :} The evolution of the energy density contours
from the expansion of two adjacent Gaussian hot spots
is shown for the case of an ideal gas equation of state.
Note the development of two back-to-back transverse jets from the
shock zone.
The time step between different frames is $0.8$ fm/c, the initial time
is 0.5 fm/c.
\\ ~~ \\
{\bf Fig.\ 9 :} Azimuthally asymmetric 
transverse shocks produced during the evolution
of the inhomogeneous initial condition in Fig.\ 8 
are evident as spikes in the transverse energy distribution (upper panels) as 
well as enhanced $E_\perp$ correlations (lower panels). 
The initial ${\rm d}E_\perp/{\rm d}y {\rm d}\phi$ and $C_{ET}$ are
straight lines, due to the absence of collective motion
in the initial conditions. Note that a first order transition (right panels)
inhibits the formation of such transverse shocks 
because the expansion is stalled
in that case (see Fig.\ 10). In both cases, the dynamically produced
asymmetries are much larger than the numerical errors from Fig.\ 7.
\\ ~~ \\
{\bf Fig.\ 10:} The evolution of the same initial condition as in Fig.\ 8
but assuming here the Bag model equation of state.
The stalled expansion due to the vanishing of the speed of sound
is revealed using larger time steps ($1.6$ fm/c) between frames relative 
to Fig.\ 8.
Note that the expansion bubbles are much thinner as is the shock zone.
\\ ~~ \\
{\bf Fig.\ 11:} The evolution of the energy density ${\cal E}$
along the $x-$axis of Figs.\ 8 and 9 shows clearly the difference between
shell structures formed in the ideal gas vs.\ the first order
transition cases. Lower graphs show the evolution of the $v^x(x,0)$
transverse velocity field that vanishes initially.
The time steps shown for the ideal case are $\tau=0.5,2.1,3.7,5.3,7.7$ fm/c,
while in the first order transition case 
they are $\tau=0.5,3.7,6.9,10.1,14.9$ fm/c. 
\\ ~~ \\
{\bf Fig.\ 12:} The evolution of the energy density ${\cal E}$
for a typical HIJING event assuming an ideal gas equation state.
The (solid) contour indicates the hadronization surface,
where the proper energy density has the value $\epsilon_H\simeq 0.1\; \gfm$.
The time steps are indicated on the top right of each frame.
Darker shaded regions correspond to hot spots while lighter shaded regions
to cooler domains. See Fig.\ 14 below for a slice along the $x-$axis
through these profiles that indicates the energy density scale.
\\ ~~ \\
{\bf Fig.\ 13:} The evolution of the same HIJING event as in Fig.\ 12, but with
the equation of state with a first order transition. 
Note that the time steps are twice as large
as in Fig.\ 12 due to the stalled expansion with this equation of state.
Also note the formation of thin bubble walls 
of mixed phase matter as in Fig.\ 10.
The contour of $\epsilon=\epsilon_H$ shows that hadronic
foam-like structures develop due to the initial state inhomogeneities. 
\\ ~~ \\
{\bf Fig.\ 14:} A slice along the $x-$axis
through selected frames in Figs.\ 12 and 13
indicates the magnitude of the energy densities in the structures formed
during evolution. The $v^x(x,0)$ collective flow velocity field
does not vanish at $\tau_{th}$ as in Fig.\ 11 because of the initial
turbulence induced by mini-jet formation. The curves are plotted at
the same times as in Fig.\ 11.
\\ ~~ \\
{\bf Fig.\ 15:} The evolution of the thermal smeared transverse energy
distribution corresponding to the event in Figs.\ 12 and 13.
Note that the initial azimuthal anisotropy reflects the turbulent
initial condition. At freeze-out, multiple narrow spikes formed 
from directed transverse
shocks (volcanoes) are superimposed 
on top of the broader initial state anisotropy.
The overall magnitude of the transverse energy is reduced 
more in the ideal case.
\\ ~~ \\
{\bf Fig.\ 16:} The event-averaged transverse energy distribution 
(upper panels) is
approximately rotation invariant within the finite statistics.
However, a clear enhancement of the transverse energy correlations is observed
(lower panels).
The initial state correlations (dashed curves)
are typically less than 1\% and the final dynamical
correlation due to transverse shocks (solid curves)
are strongest in the ideal case (lower left panel). 
\\ ~~ \\
{\bf Fig.\ 17:} The dependence of the $E_\perp$ correlation function
on the soft beam-jet background and the mini-jet $p_0$ scale is shown.
The solid curves correspond to the default HIJING $p_0=2$ GeV/c and 
$E_\perp^{soft}\equiv
{\rm d}E_\perp^{soft}/{\rm d}y=0.5$ TeV shown in Fig.\ 16. 
For the ideal gas case (left panels) 
the dashed and dotted curves correspond to 
$p_0=2$ GeV/c and $E_\perp^{soft}=0.25,1.0$ TeV, 
respectively. For the first order transition case (right panels) 
the dashed curves correspond to 
$p_0=2$ GeV/c and $E_\perp^{soft}=0.25$ TeV and the dotted curves to 
$p_0=1$ GeV/c and $E_\perp^{soft}=1.0$ TeV, 
respectively. The dash--dotted curve corresponds to $p_0=1$ GeV/c and 
$E_\perp^{soft}=0.5$ TeV. 
The lower panels show the ratio of the final to initial
correlation functions in the restricted $\Delta\phi<\pi/2$ domain.
\\ ~~ \\
{\bf Fig.\ 18:} The evolution of the central core $(r_\perp<3$ fm) mean energy
density $\langle {\cal E} \rangle $
is compared for 3+1--dimensional (thick solid and dashed curves)
and one--dimensional 
Bjorken expansion (thin dashed and dash--dotted curves)
and different equations of state (thick solid: first order transition,
thick dashed and thin dash--dotted: ideal gas, thin dashed: $p=0$).
\\ ~~ \\
{\bf Fig.\ 19:} The evolution of the mean and rms fluctuations of the central
core energy density $\langle {\cal E} \rangle$ 
is shown for the first order transition case.
The dashed curve is the same as in Fig.\ 18.
\\ ~~ \\
{\bf Fig.\ 20:} Photon transverse momentum spectra at $y=\eta = 0$.
Thick lines are the average spectra calculated from the hydrodynamical
evolution of 10 (turbulent) HIJING
events, thin lines are for the evolution of 
a homogeneous Bjorken cylinder of radius
$R=4$ fm with the same initial energy content. 
Dashed lines are obtained assuming
the equation of state with phase transition, full lines are for the ideal
gas equation of state.

\end{document}